\documentclass{IEEEtran}
\usepackage{amssymb}
\usepackage{amsfonts}
\usepackage{amsmath}
\usepackage{algorithm}
\usepackage{algorithmic}
\usepackage{tikz}
\usepackage{graphicx,subfigure,cite,multicol,multirow,diagbox,booktabs,array}
\usepackage{color}

\ifCLASSINFOpdf
\else
\fi

\begin{document}
%
\title{ Energy-efficient Alternating Iterative Secure Structure of Maximizing Secrecy Rate for Directional Modulation Networks \\ }

\author{Linlin Sun,
        Jiayu~Li,~Yu~Zhang,~Yuntian~Wang,~Linqing~Gui,~Fanyuan~Li,~Haochen~Li,
        \\~and~Zhihong~Zhuang,~Feng~Shu

\thanks{This work was supported in part by the National Natural Science Foundation of China (Nos. 61771244, 61472190, 61501238, and 61702258) \emph{(Corresponding authors: Zhihong Zhuang, Linqing Gui and Feng Shu)}.}
\thanks{Linlin Sun, Jiayu~Li,~Yuntian~Wang,~Linqing~Gui,~Fanyuan~Li,~Zhihong~Zhuang~and~Feng Shu are with the School of Electronic and Optical Engineering, Nanjing University of Science and Technology, 210094, China.}
\thanks{Yu~Zhang is with Nanjing Research Institute of Electronics Technology.}
\thanks{Haochen~Li is with Southeast University, 211189, China.}
}
\maketitle

\begin{abstract}
In a directional modulation (DM) network, the issues of security and privacy have taken on an increasingly important role. Since the power allocation of confidential message and artificial noise will make a constructive effect on the system performance, it is important to jointly consider the relationship between the beamforming vectors and the power allocation (PA) factors.
To maximize the secrecy rate (SR), an alternating iterative structure (AIS) between the beamforming and PA is proposed. With only two or three iterations, it can rapidly converge to its rate ceil. Simulation results indicate that the SR performance of proposed AIS is much better than the null-space projection (NSP) based PA strategy in the medium and large signal-to-noise ratio (SNR) regions, especially when the number of antennas at the DM transmitter is small.
\end{abstract}

\maketitle

\section{Introduction}
\IEEEPARstart{I}{n} recent years, security and privacy of confidential information increasingly becomes an extremely important problem in wireless networks and next generation wireless systems\cite{WH-DBPHY,ZN-AEIA,CX-SMAPHY,WY-STLA}. Techniques such as orthogonal frequency division multiplexing (OFDM), massive multiple-input multiple-output (MIMO) and hybrid beamforming have been introduced in future fifth generation (5G) cellular systems, internet of things (IoT), unmanned aerial vehicle (UAV), smart transportation, and satellite communications
\cite{WANG-OFDMA1,WANG-OFDMA2,massivemimo,KG-5Ghspeed1,KG-5Ghspeed2,hbsurvey}.

Directional modulation (DM), as a green and efficient secure transmission scheme, offers security through its directive property and is suitable for line-of-propagation (LoP) channels.
Some energy efficient techniques have been taken into consideration in green communication to strike a good balance between throughput and energy consumption\cite{ZH-SWIPT,WuQ-viewGreen5G,MMGreen5G}.
DM can perform effective beam alignment and information synthesis in the RF portion, and then concentrate the energy in the direction of the intended receiver, which facilitates the energy harvesting at the receiver.
The great potential of DM has been highlighted as a key enabling secure technology for the next generation wireless systems and green communication.
As the number of transmit antennas tends to medium-scale or large-scale, the energy of confidential message (CM) and artificial noise (AN) is transmitted via the corresponding narrow beams with little energy wasting, thus DM can implement an energy-efficient transmission from this aspect.

As the concept of secrecy capacity was proposed for a discrete memoryless wiretap channel in \cite{Wyner1975}, AN was utilized in \cite{Goel,YD-avector,ANYang,AN-wiretap-YSH} to enhance the information-theoretic security.
The authors in \cite{HU2016-RDM} first introduced a robust DM synthesis method of null-space projection (NSP) to project the AN along the eavesdropping direction on the null space of the steering vector. \cite{WU2016-RDMBC}, \cite{MU-MIMO-ZHU} and \cite{DM-XU} proposed several synthesis methods and secure schemes for three different scenarios: multi-user broadcasting, multi-user MIMO and multicast DM scenario to improve the security.
Furthermore, a practical DM scheme with random frequency diverse array with the aid of AN is proposed to enhance physical layer security \cite{RFDA-HJS}. A new concept of secure and precise wireless transmission (SPWT) is proposed to achieve SPWT of CM in \cite{Wu-SPTDM}, which combined the techniques of AN projection, beamforming and random subcarrier selection. Since direction-of-arrival (DOA) scheme is of great significance for practical DM applications, the authors in \cite{DOA-QIN} proposed three estimators of DOA based on hybrid structure to determine the position, which makes DM more feasible.

With the increasing demand for secure transmission in DM systems, power allocation (PA) now becomes particularly important. In \cite{Xiangyun,Tsai,Yongpeng,TongXing,Huanhuan,WAN-PA,LU-PA,LU-PA2}, the authors proposed several strategies and analysis of optimal power allocation (OPA) in different situations. The authors in \cite{Xiangyun} studied the OPA strategy in the presence of noncolluding or colluding eavesdroppers. \cite{Tsai} and \cite{Yongpeng} investigated secure and reliable transmission strategies for MIMO systems and obtained OPA policy for the transmit signal and AN. Since secrecy rate (SR) is an important factor in secure communication, authors in \cite{TongXing} and \cite{Huanhuan} derived the optimal solutions of PA by maximizing SR.
In \cite{WAN-PA}, the closed-form expression of OPA factors can be derived by using the Lagrange multiplier method, with determined beamforming vector and AN projection matrix based on NSP scheme. The authors in \cite{LU-PA} and \cite{LU-PA2} proposed several OPA methods in secure DM networks with UAV users.

Alternating iterative scheme can transform the multivariate problem into a univariate problem. The basic idea is to keep the other quantities fixed and calculate only one variable in each step of the calculation process. \cite{LU-PA} proposed an OPA strategy based on maximizing signal-to-leakage-and-noise ratio (Max-SLNR) and maximizing AN-and-leakage-to-noise ratio (Max-ANLNR) to design beamforming vector and AN projection vector, so as to maximize SR (Max-SR).
This AIS between PA and beamforming can improve the average SR in the UAV networks.
However, there is no relative study to design both the beamforming vector and AN projection vector using Max-SR method.

In this paper, we first propose an alternating iterative secure structure of Max-SR for DM networks. We will use the method of Max-SR scheme with combination of general power iterative (GPI) based beamforming scheme in \cite{GPI-HY} to design and optimize the beamforming vector, AN projection vector and PA factors alternatively.
To simplify the iterative steps, we initialize the AN projection vector using Max-ANLNR method, and initialize PA factors in a fixed value. Then, we maximize SR to get the beamforming vector. Next, using the designed beamforming vector and initialized AN projection vector, the PA factors are computed based on Max-SR strategy. Then in each iteration, we design the beamforming vector, AN projection vector and PA factors one by one based on {Max-SR} method. This process is repeated until the terminal condition is satisfied. Meanwhile the Max-SR result of our proposed AIS method can be obtained.

The remainder of this paper is organized as follows. Section \ref{S2} describes the system model. In Section \ref{S3}, the beamforming vector and AN projection vector are given, and the AIS of Max-SR is proposed. Simulation results are presented in Section \ref{S4}. Finally, we make our conclusions in Section \ref{S5}.

\emph{Notations:} throughout the paper, matrices, vectors, and scalars are denoted by letters of bold upper case, bold lower case, and lower case, respectively. Signs $(\cdot)^T$, $(\cdot)^*$, $(\cdot)^H$ and $|\cdot|$ denote transpose, conjugate, conjugate transpose and modulus respectively. Notation $\mathbb{E}\{\cdot\}$ stands for the expectation operation. $\textbf{I}_N$ denotes the $N\times N$ identity matrix.

\section{System Model}\label{S2}
\begin{figure}[htb]
  \centering
  \includegraphics[width=0.45\textwidth]{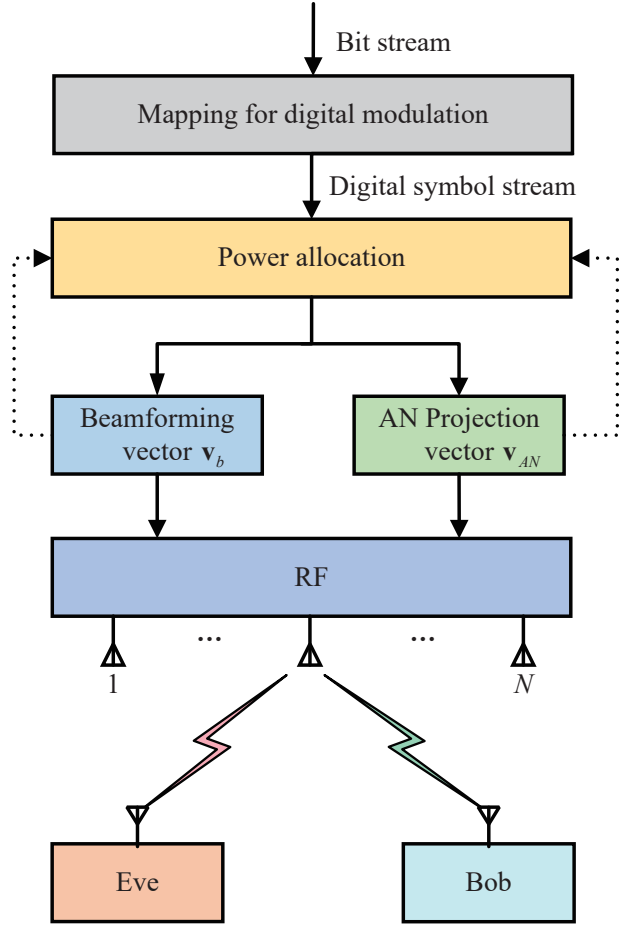}
  \caption{Schematic diagram of the proposed directional modulation system.}
  \label{sys}
\end{figure}
As shown in Fig. \ref{sys}, we consider a DM system, where Alice is equipped with $N$ antennas, Bob and Eve are equipped with single antenna, respectively. In this paper, we assume there exists the line-of-sight (LOS) path. The transmitted baseband signal can be expressed as
\begin{equation}\label{Tx signal s}
\mathbf{s}=\sqrt{\beta P_s}\mathbf{v}_bx+\sqrt{(1-\beta)P_s}\mathbf{v}_{AN}z,
\end{equation}
where $P_s$ is the total transmission power and limited, $\beta$ and  $(1-\beta)$ are the PA parameters of CM and AN, respectively. $\mathbf{v}_b\in\mathbb{C}^{N\times1}$ denotes the transmit beamforming vector for controlling the CM to the desired direction and $\mathbf{v}_{AN}\in\mathbb{C}^{N\times 1} $ is the projection vector leading AN to the undesired direction, where $\mathbf{v}_b^H\mathbf{v}_b=1$ and $\mathbf{v}_{AN}^H\mathbf{v}_{AN}=1$. In (\ref{Tx signal s}),  $x$ is the CM of satisfying $\mathbb{E}\left\{x^Hx\right\}=1$ and $z\sim\mathcal{CN}(0,\mathbf{I}_N)$ denotes the scalar AN being a complex Gaussian random variable with zero mean and unit variance.

The received signal at Bob is given by
\begin{equation}\label{eq2}
\begin{aligned}
y\left(\theta_{b}\right) &=\mathbf{h}^{H}\left(\theta_{b}\right) \mathbf{s}+n_{b}\\
&=\sqrt{\beta P_{s}} \mathbf{h}^{H}\left(\theta_{b}\right) \mathbf{v}_{b} {x}\\
&+\sqrt{(1-\beta) P_{s}} \mathbf{h}^{H}\left(\theta_{b}\right) \mathbf{v}_{A N} {z}+n_{b},
\end{aligned}	
\end{equation}
where $\mathbf{h}(\theta_b)\in\mathbb{C}^{N\times1}$ represent the channel vector between Alice and Bob, and $n_b\sim\mathcal{C}\mathcal{N}(0,\sigma_b^2)$ denotes the complex additive white Gaussian noise (AWGN) at Bob. Similarly, the received signal at Eve can be written as
\begin{equation}\label{eq3}
\begin{aligned}
y\left(\theta_{e}\right) &=\mathbf{h}^{H}\left(\theta_{e}\right) \mathbf{s}+n_{e} \\
&=\sqrt{\beta P_{s}} \mathbf{h}^{H}\left(\theta_{e}\right) \mathbf{v}_{b} {x}\\
&+\sqrt{(1-\beta) P_{s}} \mathbf{h}^{H}\left(\theta_{e}\right) \mathbf{v}_{A N} {z}+n_{e},
\end{aligned}
\end{equation}
where $\mathbf{h}(\theta_e)\in\mathbb{C}^{N\times1}$ represent the channel vector between Alice and Eve, and $n_e\sim\mathcal{C}\mathcal{N}(0,\sigma_e^2)$ denotes the complex additive white Gaussian noise (AWGN) at Eve. In the following, we assume that $\sigma_b^2=\sigma_e^2=\sigma^2$.

According to (\ref{eq2}) and (\ref{eq3}), the achievable rates from Alice to Bob and to Eve  can be expressed as
\begin{equation}\label{eq4}
R\left(\theta_{b}\right)=\log _{2}\left(1+\frac{\beta P_{s}\left|\mathbf{h}^{H}\left(\theta_{b}\right) \mathbf{v}_{b}\right|^{2}}{(1-\beta) P_{s}\left|\mathbf{h}^{H}\left(\theta_{b}\right) \mathbf{v}_{A N}\right|^{2}+\sigma^{2}}\right),
\end{equation}
and
\begin{equation}\label{eq5}
R\left(\theta_{e}\right)=\log _{2}\left(1+\frac{\beta P_{s}\left|\mathbf{h}^{H}\left(\theta_{e}\right) \mathbf{v}_{b}\right|^{2}}{(1-\beta) P_{s}\left|\mathbf{h}^{H}\left(\theta_{e}\right) \mathbf{v}_{A N}\right|^{2}+\sigma^{2}}\right),
\end{equation}
respectively. As a result, the achievable secrecy rate can be written as
\begin{equation}\label{eq6}
R_{s}=\max \left\{0, R\left(\theta_{b}\right)-R\left(\theta_{e}\right)\right\}.
\end{equation}

\section{Proposed AIS Max-SR based PA scheme }\label{S3}
According to (\ref{eq4}), (\ref{eq5}) and (\ref{eq6}), the secrecy rate is a function with respect to $\beta$, $\mathbf{v}_{AN}$ and $\mathbf{v}_b$. Consequently, the optimization problem of maximizing the secrecy rate can be formulated as
\begin{subequations}\label{P1}
\begin{equation}
\underset{\mathbf{v}_{b},\mathbf{v}_{A N},\beta}{\max}~R_s\left(\beta,\mathbf{v}_{AN},\mathbf{v}_{b}\right)
\end{equation}
\begin{equation}
\text{s.t.}~~0 \leqslant \beta \leqslant 1
\end{equation}
\begin{equation}
~~~~~~~~~\mathbf{v}_{A N}^{H} \mathbf{v}_{A N}=1
\end{equation}
\begin{equation}
~~~~~\mathbf{v}_{b}^{H} \mathbf{v}_{b}=1
\end{equation}
\end{subequations}
where $R_s(\beta,\mathbf{v}_{A N},\mathbf{v}_{b})$ represents the secrecy rate and is given by
\begin{equation}
\begin{aligned}
&R_s\left(\beta,\mathbf{v}_{A N},\mathbf{v}_{b}\right)=R(\theta_{b})-R(\theta_{e})\\
&=\log_{2}\Bigg[\left(\frac{(1\!-\!\beta) P_{s}\left|\mathbf{h}^{H}\left(\theta_{b}\right) \mathbf{v}_{A N}\right|^{2}\!+\!\sigma^{2}\!+\!\beta P_{s}\left|\mathbf{h}^{H}\left(\theta_{b}\right) \mathbf{v}_{b}\right|^{2}}{(1\!-\!\beta) P_{s}\left|\mathbf{h}^{H}\left(\theta_{b}\right) \mathbf{v}_{A N}\right|^{2}\!+\!\sigma^{2}}\right)\\
&\times\left(\frac{(1\!-\!\beta)P_{s}\left|\mathbf{h}^{H}\left(\theta_{e}\right) \mathbf{v}_{A N}\right|^{2}\!+\!\sigma^{2}}{(1\!-\!\beta) P_{s}\left|\mathbf{h}^{H}\left(\theta_{e}\right)\mathbf{v}_{A N}\right|^{2}\!+\!\sigma^{2}\!+\!\beta P_{s}\left|\mathbf{h}^{H}\left(\theta_{e}\right)\mathbf{v}_{b}\right|^{2}}\right)\Bigg].
\end{aligned}
\end{equation}

Obviously, it is difficult to solve the joint optimization problem (\ref{P1}). Therefore, we propose an iterative algorithm in the following. Specifically, for given $\mathbf{v}_{AN}$ and $\beta$, we design $\mathbf{v}_b$. Then, for given $\mathbf{v}_b$ and $\beta$, we design $\mathbf{v}_{AN}$. Finally, for given $\mathbf{v}_{AN}$ and $\mathbf{v}_b$, we design $\beta$.

Observing (\ref{Tx signal s}), we find the fact that the sum of CM and AN power is equal to $P_s$, this means that the optimization problem of Max-SR in problem (\ref{P1}) is addressed with implicit power constraint. According to the definition of secure energy efficiency (EE), i.e. $\frac{R_s}{P_s}$, it is obvious that maximizing the value of $R_s$ can achieve a high secure EE value of $\frac{R_s}{P_s}$ for a fixed value of $P_s$. In other words, given a fixed value of $P_s$, the optimization problem of Max-SR in problem (\ref{P1}) can accomplish a high secure EE.

\subsection{Beamforming Vector Optimization}
For any given $\mathbf{v}_{AN}$ and $\beta$, problem (\ref{P1}) can be simplified as
\begin{subequations}
\begin{equation}
\underset{\mathbf{v}_b}{\max}~~R_s(\mathbf{v}_{b})
\end{equation}
\begin{equation}
~~~~~~\text {s.t.}~~\mathrm{v}_{b}^{H}\mathrm{v}_{b}=1
\end{equation}
\end{subequations}
where
\begin{equation}
A=(1-\beta )\left|\mathbf{h}^{H} (\theta_{b})\mathbf{v}_{A N}\right|^{2}+\frac{\sigma^{2}}{P_{s}},
\end{equation}
\begin{equation}
B=(1-\beta )\left|\mathbf{h}^{H} (\theta_{e})\mathbf{v}_{A N}\right|^{2}+\frac{\sigma^{2}}{P_{s}},
\end{equation}
\begin{equation}
R_s\left(\mathbf{v}_{b}\right)=\log _{2}\left(\frac{AB+\beta B\left|\mathbf{h}^{H}\left(\theta_{b}\right) \mathbf{v}_{b}\right|^{2}}{AB+\beta A\left|\mathbf{h}^{H}\left(\theta_{e}\right) \mathbf{v}_{b}\right|^{2}}\right).
\end{equation}
According to the Rayleigh-Ritz theorem, the optimal $\mathbf{v}_b$ can be obtained from the eigenvector corresponding to the largest eigenvalue of the matrix
\begin{equation}
\left[AB \mathbf{I}_{N}\!+\!\beta A\mathbf{h}\left(\theta_{e}\right) \mathbf{h}^{H}\left(\theta_{e}\right)\right]^{-1} \left[AB\mathbf{I}_{N}\!+\!\beta B\mathbf{h}\left(\theta_{b}\right) \mathbf{h}^{H}\left(\theta_{b}\right)\right].
\end{equation}

\subsection{AN Projection Vector Optimization}
For any given $\mathbf{v}_{b}$ and $\beta$, problem (\ref{P1}) can be reformulated as
\begin{subequations}
\begin{equation}
\underset{\mathbf{v}_{AN}}{\max}~~R_s(\mathbf{v}_{AN})
\end{equation}
\begin{equation}
~~~~~~~~~\text {s.t.}~~\mathrm{v}_{AN}^{H}\mathrm{v}_{AN}=1
\end{equation}
\end{subequations}
where
\begin{equation}
\mathbf{C}=\big(\frac{\beta}{1-\beta}\left|\mathbf{h}^{H}\left(\theta_{b}\right) \mathbf{v}_{b}\right|^{2}+\frac{\sigma^{2}}{(1-\beta)P_{s}}\big)
\mathbf{I}_N+\mathbf{h}(\theta_b)\mathbf{h}^H(\theta_b),
\end{equation}
\begin{equation}
\mathbf{D}=\big(\frac{\beta}{1-\beta}\left|\mathbf{h}^{H}\left(\theta_{e}\right) \mathbf{v}_{b}\right|^{2}+\frac{\sigma^{2}}{(1-\beta)P_{s}}\big)
\mathbf{I}_N+\mathbf{h}(\theta_e)\mathbf{h}^H(\theta_e),
\end{equation}
\begin{equation}
\mathbf{E}=\big(\frac{\sigma^{2}}{(1-\beta)P_{s}}\big)\mathbf{I}_N+
\mathbf{h}(\theta_e)\mathbf{h}^H(\theta_e),
\end{equation}
\begin{equation}
\mathbf{F}=\big(\frac{\sigma^{2}}{(1-\beta)P_{s}}\big)\mathbf{I}_N+
\mathbf{h}(\theta_b)\mathbf{h}^H(\theta_b),
\end{equation}
\begin{equation}\label{eqrsvan}
R_s\left(\mathbf{v}_{AN}\right)=\log_{2}\left(\frac{\mathbf{v}_{AN}^{H}\mathbf{C}\mathbf{v}_{AN}}
{\mathbf{v}_{AN}^{H}\mathbf{D}\mathbf{v}_{AN}}\times
\frac{\mathbf{v}_{AN}^{H}\mathbf{E}\mathbf{v}_{AN}}{\mathbf{v}_{AN}^{H}\mathbf{F}\mathbf{v}_{AN}} \right).
\end{equation}
Since (\ref{eqrsvan}) is a non-convex quadratic fractional function and $\mathbf{C}$, $\mathbf{D}$, $\mathbf{E}$, $\mathbf{F}$ are positive semi-definite matrices, the optimal $\mathbf{v}_{AN}$ can be obtained by utilizing the GPI algorithm\cite{GPI-HY}.

\subsection{Power Allocation Parameter Optimization}
For any given $\mathbf{v}_{AN}$ and $\mathbf{v}_{b}$, problem (\ref{P1}) can be rewritten as
\begin{subequations}
\begin{equation}
\underset{\beta}{\max}~~R_s(\beta)
\end{equation}
\begin{equation}
~~~~~~~~\text {s.t.}~~0 \leqslant \beta \leqslant 1
\end{equation}
\end{subequations}
where
\begin{equation}\label{eq15}
\begin{aligned}
R_s(\beta)
&=\log_2\frac{\overbrace{I\beta^2+J\beta+K}^{a(\beta)}}{\underbrace{L\beta^2+M\beta+K}_{b(\beta)}}\\
&=\log_2\frac{a(\beta)}{b(\beta)}\\
&=\log_2\phi(\beta),
\end{aligned}
\end{equation}
\begin{align}
I&=P_s^2 |\mathbf{h}^H(\theta_e)\mathbf{v}_{AN}|^2\nonumber\\
&\times\left(|\mathbf{h}^H(\theta_b)\mathbf{v}_{AN}|^2-|\mathbf{h}^H(\theta_b)\mathbf{v}_b|^2 \right),\\
J&=P_s\left(|\mathbf{h}^H(\theta_b)\mathbf{v}_b|^2-|\mathbf{h}^H(\theta_b)\mathbf{v}_{AN}|^2\right)\nonumber\\
&\times\left(P_s |\mathbf{h}^H(\theta_e)\mathbf{v}_{AN}|^2+\sigma^2\right)\nonumber\\
&-P_s |\mathbf{h}^H(\theta_e)\mathbf{v}_{AN}|^2
\left(P_s |\mathbf{h}^H(\theta_b)\mathbf{v}_{AN}|^2+\sigma^2\right),\\
K&=\left(P_s |\mathbf{h}^H(\theta_b)\mathbf{v}_{AN}|^2+\sigma^2\right)\nonumber\\
&\times\left(P_s |\mathbf{h}^H(\theta_e)\mathbf{v}_{AN}|^2+\sigma^2\right),\\
L&=P_s^2 |\mathbf{h}^H(\theta_b)\mathbf{v}_{AN}|^2\nonumber\\
&\times\left(|\mathbf{h}^H(\theta_e)\mathbf{v}_{AN}|^2-|\mathbf{h}^H(\theta_e)\mathbf{v}_b|^2 \right),\\
M&=P_s\left(|\mathbf{h}^H(\theta_e)\mathbf{v}_b|^2-|\mathbf{h}^H(\theta_e)\mathbf{v}_{AN}|^2\right)\nonumber\\
&\times\left(P_s|\mathbf{h}^H(\theta_b)\mathbf{v}_{AN}|^2+\sigma^2\right)\nonumber\\
&-P_s |\mathbf{h}^H(\theta_b)\mathbf{v}_{AN}|^2
\left(P_s |\mathbf{h}^H(\theta_e)\mathbf{v}_{AN}|^2+\sigma^2\right).
\end{align}
Under the total transmit power constraint, the secrecy rate given by (\ref{eq15}) is also limited. This means that $a(\beta)$ and $b(\beta)$ in (\ref{eq15}) should be not equal to zero. Otherwise, an infinite value of secrecy rate is generated. As a result, $\phi(\beta)$ should be not equal to zero as well. To maximize the secrecy rate, let the derivative of $R_s(\beta)$ with respect to $\beta$ be equal to zero, which yields
\begin{equation}\label{eq22}
\frac{\partial R_s(\beta)}{\partial \beta}=\frac{1}{\phi(\beta)} \frac{\partial \phi(\beta)}{\partial \beta}=0.
\end{equation}
Note that $\phi(\beta)\neq 0$, then (\ref{eq22}) is equivalent to
\begin{equation}\label{eq23}
\frac{\partial\phi(\beta)}{\partial\beta}=\frac{(IM\!-\!JL)\beta^2+2K(I\!-\!L)\beta+K(J\!-\!M)}{(L\beta^2+M\beta+K)^2}=0.
\end{equation}
Considering the fact that $(L\beta^2+M\beta+K)^2\neq 0$, (\ref{eq23}) is equivalent to
\begin{equation}\label{eq24}
(IM-JL)\beta^2+2K(I-L)\beta+K(J-M) = 0.
\end{equation}
If $IM-JL\neq 0$, (\ref{eq24}) is a quadric equation and $\Delta=K^2(I-L)^2-K(IM-JL)(J-M)$. If $\Delta\geq 0$, the singular points $\beta_1$ and $\beta_2$ are given by
\begin{equation}
\beta_{1}=\frac{-K(I-L)+\sqrt{\Delta}}{IM-JL},
\end{equation}
\begin{equation}
\beta_{2}=\frac{-K(I-L)-\sqrt{\Delta}}{IM-JL},
\end{equation}
respectively. If $IM-JL=0$, (\ref{eq24}) is a linear equation and the singular point $\beta_3$ is given by
\begin{equation}
\beta_{3}=\frac{M-J}{2(I-L)}.
\end{equation}

In what follows, we need to determine whether these singular points are in the interval $(0,1)$, and then obtain the optimal power allocation parameter. Based on the above derivation, the power allocation strategy is concluded in Algorithm \ref{algorithm 1}.
\begin{algorithm}
\caption{Proposed optimal power allocation strategy}\label{algorithm 1}
\begin{algorithmic}[1]
\STATE Initialize $P_s$, $\theta_b$, $\theta_e$, $\mathbf{h}(\theta_b)$, $\mathbf{h}(\theta_e)$, $\mathbf{v}_b$, $\mathbf{v}_{AN}$ and SNR.
\IF{$IM-JL=0$}
\STATE \textbf{Case 1.} If $\beta_{3} \in(0,1)$, then the optimal
power allocation parameter $\beta^*=\mathop{\arg\max}\{\phi(0),\phi(\beta_3),\phi(1)\}$.
\STATE \textbf{Case 2.} If $\beta_{3} \notin(0,1)$, then $\beta^*=\mathop{\arg\max}\{\phi(0),\phi(1)\}$.
\ELSIF{$IM-JL\neq0$ and $\Delta\geq0$}
\STATE \textbf{Case 1.} If $\beta_1\in(0,1)$, $\beta_2\in(0,1)$, then  $\beta^*=\mathop{\arg\max}\{\phi(0),\phi(\beta_1),\phi(\beta_2),\phi(1)\}$.\\
\STATE \textbf{Case 2.} If $\beta_1\in(0,1)$, $\beta_2\notin(0,1)$, then $\beta^*=\mathop{\arg\max}\{\phi(0),\phi(\beta_1),\phi(1)\}$.\\
\STATE \textbf{Case 3.} If $\beta_1\notin(0,1)$, $\beta_2\in(0,1)$, then $\beta^*=\mathop{\arg\max}\{\phi(0),\phi(\beta_2),\phi(1)\}$.\\
\STATE \textbf{Case 4.} If $\beta_1\notin(0,1)$, $\beta_2\notin(0,1)$, then $\beta^*=\mathop{\arg\max}\{\phi(0),\phi(1)\}$.
\ELSE
\STATE \textbf{Case 1.} If $IM-JL>0$, then $\phi(\beta)$ is a monotonically increasing function of $\beta$. Therefore, $\beta^*=1$.\\
\STATE \textbf{Case 2.} If $IM-JL<0$, then $\phi(\beta)$ is a monotonically decreasing function of $\beta$. Therefore, $\beta^*=0$.
\ENDIF
\end{algorithmic}
\end{algorithm}

There are two special cases in the above power allocation strategy, which are detailedly discussed in the following.
\begin{itemize}
\item When $\beta^*=1$, all power of Alice is used to transmit CM, and AN fails to work. At this time, the secrecy rate $R_s^*=R_s(1)=\log_2(\frac{I+J+K}{L+M+K})$, and the DM system degenerates into a general multi-antenna transmitter. In general, this case is likely to exist in two scenarios:
    \\(1) The transmitter is equipped with a lot of antennas, and consequently the transmission beam is narrow.
    \\(2) The quality of the transmission signal is poor, and SNR is poor as well.
\item When $\beta^*=0$,  all power of Alice is used to transmit AN, and no CM is transmitted to Bob. At this time, the secrecy rate $R_s^*=R_s(0)=0$, and secure communication cannot be guaranteed. Therefore, this case should be avoided.
\end{itemize}

\subsection{Overall Algorithm}
\begin{figure}[htb]
  \centering
  \includegraphics[width=0.45\textwidth]{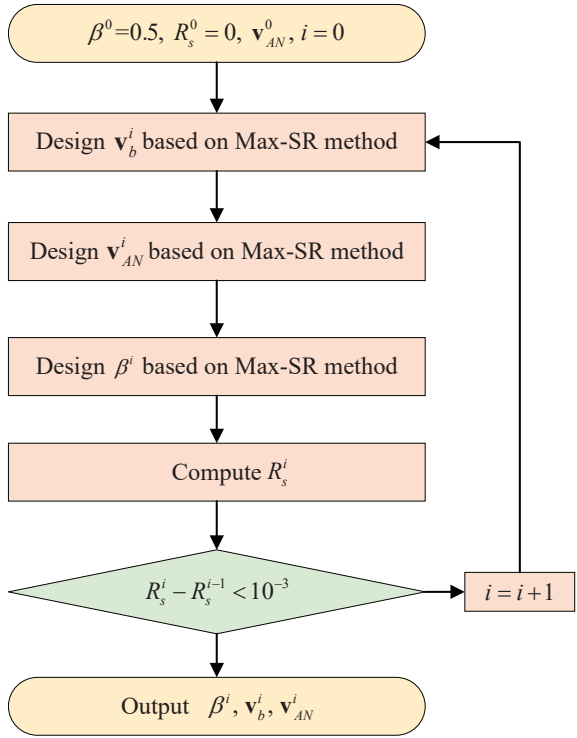}
  \caption{Proposed iterative algorithm for problem (\ref{P1}).}
  \label{liutu}
\end{figure}

Based on the results in the previous three subsections, we proposed an iterative algorithm for problem (\ref{P1}), which is summarized in Algorithm \ref{A1}. To make it clear, the detailed procedure is also indicated in Fig. \ref{liutu}. Specifically, we first initialize $\beta^0=0.5$, $R_s^0=0$, $i=1$ and design $\mathbf{v}_{AN}^0$ by minimizing its leakage to Bob, called Max-ANLNR, which is formed as
\begin{subequations}
\begin{equation}
\underset{\mathbf{v}_{AN}^0}{\max}~~\text{ANLNR}(\mathbf{v}_{AN}^0, \text{fixed}~\beta^0)
\end{equation}
\begin{equation}
~~~~~~~~\text {s.t.}~~ \beta^0= 0.5
\end{equation}
\begin{equation}
~~~~~~~~~(\mathbf{v}_{A N}^0)^{H} \mathbf{v}_{A N}^0=1
\end{equation}
\end{subequations}
where
\begin{align}
&\text{ANLNR}(\mathbf{v}_{AN}^0, \text{fixed}~\beta^0)=\\\nonumber
&\frac{(\mathbf{v}_{AN}^0)^{H}[\mathbf{h}(\theta_e)\mathbf{h}^{H}(\theta_e)]
\mathbf{v}_{AN}^0}{(\mathbf{v}_{AN}^0)^{H}
[\mathbf{h}(\theta_b)\mathbf{h}^{H}(\theta_b)+\frac{\sigma^2}{(1-\beta^0)P_s)}\mathbf{I}_N]\mathbf{v}_{AN}^0}.
\end{align}
According to the Rayleigh-Ritz theorem, the optimal $\mathbf{v}_{AN}^0$ can be obtained from the eigenvector corresponding to the largest eigenvalue of the matrix
\begin{equation}
\left[\mathbf{h}(\theta_b)\mathbf{h}^{H}(\theta_b)+\frac{\sigma^2}{(1-\beta)P_s)}\mathbf{I}_N\right]^{-1}
\left[ \mathbf{h}(\theta_e)\mathbf{h}^{H}(\theta_e) \right] .
\end{equation}
Then in each iteration, we design $\mathbf{v}_{b}^i$, $\mathbf{v}_{AN}^i$ and $\beta^i$ one by one based on {Max-SR} method. Furthermore, the solution obtained in each iteration is used as the input of the next iteration. The iterations repeat until the fractional increase of $R_s$ is below $10^{-3}$.

\begin{algorithm}[htb]
\caption{Proposed iterative algorithm for problem (\ref{P1})}\label{A1}
\begin{algorithmic}[1]
\STATE Initialize $\beta^0=0.5$, $R_s^0=0$, $i=0$ and design $\mathbf{v}_{AN}^0$ based on {Max-ANLNR} method.
\REPEAT
\STATE For given $(\beta^i,\mathbf{v}_{AN}^i)$, design $\mathbf{v}_{b}^{i+1}$ based on Max-SR method.
\STATE For given $(\beta^i,\mathbf{v}_{b}^{i+1})$, design $\mathbf{v} _{AN}^{i+1}$ based on Max-SR method.
\STATE For given $(\mathbf{v}_{b}^{i+1},\mathbf{v}_{AN}^{i+1})$, design $\beta^{i+1}$ based on Max-SR method.
\STATE Update $i=i+1$ and compute $R_s^{i}$.
\UNTIL{$R_s^i-R_s^{i-1}<10^{-3}$.}
\end{algorithmic}
\end{algorithm}

\section{Simulation and Discussion}\label{S4}
\begin{figure}
  \centering
  \includegraphics[width=0.5\textwidth]{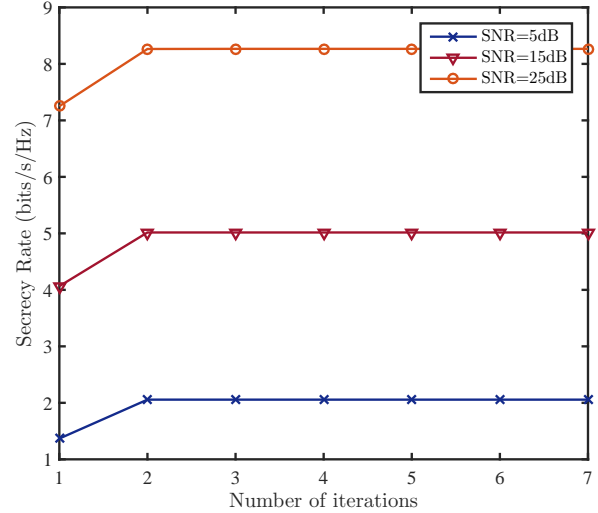}\\
  \caption{Secrecy rate versus number of iterations at $N=64$ and different transmit SNR. }\label{SR_AIS}
\end{figure}

To evaluate the SR performance gain of the proposed AIS-based Max-SR scheme, simulation results and analysis are presented in the following.

In our simulation, system parameters are set as follows: quadrature phase shift keying (QPSK) modulation, the total transmitting power $P_s=70$ dBm, the spacing between two adjacent antennas $d=\lambda/2$, the desired direction $\theta_b=45^{\circ}$, and the eavesdropping direction $\theta_e=30^{\circ}$.

Fig.~\ref{SR_AIS} plots the curves of SR versus number of iterations ranging from 1 to 7 for three typical transmit SNR : 5, 15 and 25dB, where the number of antennas at Alice is equal to 64. Observing the figure, the proposed AIS of Max-SR scheme can converge rapidly within two or three iterations. After convergence, the proposed AIS scheme may achieve an excellent SR improvement before convergence.

\begin{figure}[!htbp]
  \centering
  \subfigure[{SNR}=5dB]{
  \label {compare_NSP_AIS:SNR:a}
  \includegraphics[width=0.5\textwidth]{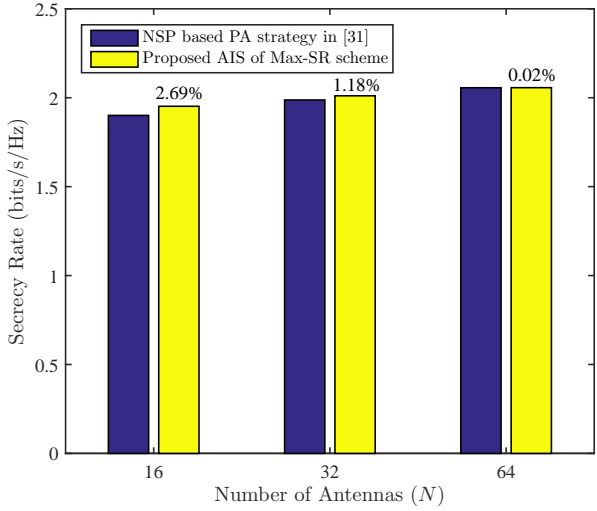}}
  \hspace{0.5in}
  \subfigure[{SNR}=15dB]{
  \label {compare_NSP_AIS:SNR:b}
  \includegraphics[width=0.5\textwidth]{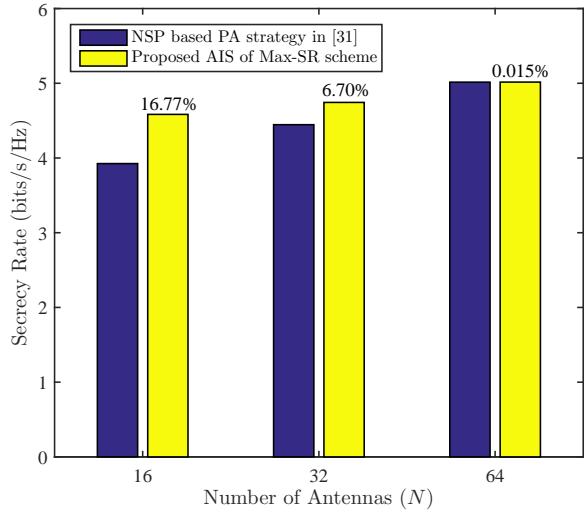}}
  \hspace{0.5in}
  \subfigure[{SNR}=25dB]{
  \label {compare_NSP_AIS:SNR:c}
  \includegraphics[width=0.5\textwidth]{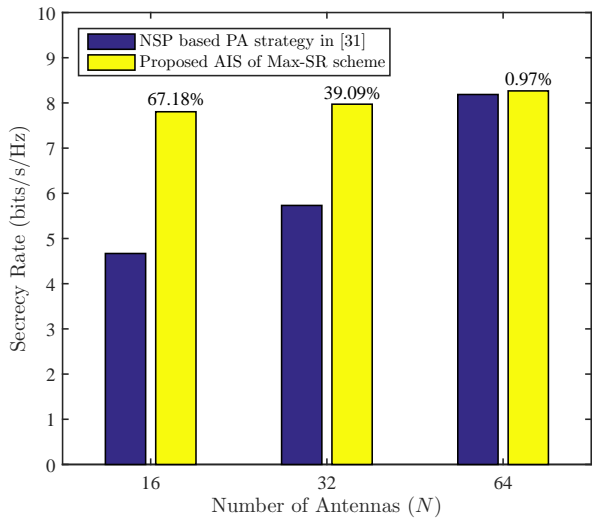}}
  \hspace{0.5in}
  \caption{Secrecy rate versus number of antennas $N$ of two different methods for different transmit SNR.}
  \label{compare_NSP_AIS:SNR}
\end{figure}

Fig.~\ref{compare_NSP_AIS:SNR} depicts the histograms of the SR versus the number of antennas at Alice $N$ for two schemes at three typical transmit SNR : 5, 15, 25dB, respectively. Compared with the NSP based PA strategy in \cite{WAN-PA}, the proposed AIS of Max-SR scheme achieves greater SR. More importantly, in the medium or high SNR, the SR performance utilizing the proposed scheme can make an improvement up to 16\% when the antenna scale is medium.

\begin{figure}
  \centering
  \includegraphics[width=0.5\textwidth]{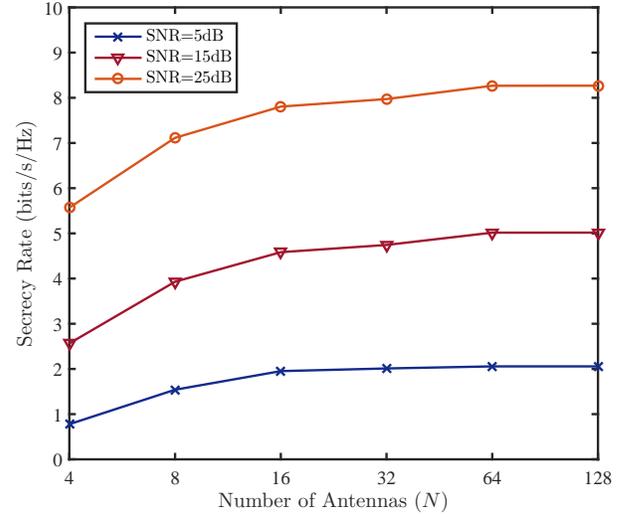}\\
  \caption{Secrecy rate versus number of antennas $N$ at different transmit SNR.}\label{SR_N}
\end{figure}

Fig.~\ref{SR_N} plots the SR versus number of antennas at three typical SNR : 5, 15, 25dB. From this figure, it is obvious that both the increase in the number of antennas and SNR can improve the SR gradually. In the case of the small number of antennas, the increase in the number of antennas can improve the SR obviously. However, for a large number of antennas, the SR performance gain achieved by doubling the number of antennas becomes smaller.

\begin{figure}
  \centering
  \includegraphics[width=0.5\textwidth]{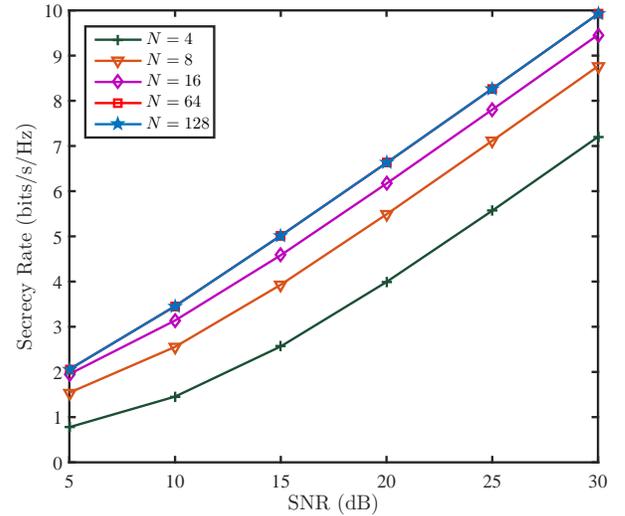}\\
  \caption{Secrecy rate versus transmit SNR at different numbers of antennas $N$.}\label{SR_SNR}
\end{figure}

Fig.~\ref{SR_SNR} demonstrates the curves of SR versus SNR with different numbers of antennas. From this figure, we can see that the curve of $N=64$ and the curve of $N=128$ almost coincide, which is in consistence with the case in Fig.~\ref{SR_N}.

\section{Conclusion}\label{S5}
In this paper, we propose an AIS to realize an iterative operation between beamforming and PA, which further improves SR based on the Max-SR scheme. Compared with the NSP based PA strategy, the proposed AIS of Max-SR scheme achieved a substantial SR improvement in the medium and large SNR regions, especially in a small or medium antenna scale. Furthermore, the proposed scheme can converge rapidly within two or three iterations and achieve an excellent SR improvement.
In the coming future, the proposed AIS of Max-SR scheme will be potentially applied to the following diverse applications such as mmWave communications, IoT systems, UAV and satellite communications.

\ifCLASSOPTIONcaptionsoff
  \newpage
\fi

\bibliographystyle{IEEEtran}
\bibliography{IEEEfull,cite}

\end{document}